\documentclass[12pt,twoside]{article}
\usepackage{amssymb}
\usepackage{amsmath}
\usepackage{latexsym}
\usepackage{longtable}
\usepackage{epsfig}
\usepackage{authblk}
\usepackage{graphicx,bbm,psfrag}
\usepackage{epsfig} 

\title{The GGE averaged currents of the classical Toda chain}
\author[1]{Xiangyu Cao}
\author[1]{Vir B. Bulchandani}
\author[2]{Herbert Spohn}
\affil[1]{Department of Phyics, University of California, Berkeley, 94720 California, United States of America.}
\affil[2]{Zentrum Mathematik and Physik Department, TUM,
	Boltzmannstr. 3, 85747 Garching, Germany.}
\begin{document}

\maketitle
\vspace{2cm}
\noindent
\textbf{Abstract}. The Toda chain with random initial data is studied. Of particular interest are generalized Gibbs ensembles, their averaged conserved fields,
and the averages of the corresponding currents. While averaged fields are well-understood, the description of averaged currents has hitherto relied on the collision-rate assumption.
For the Toda chain, the rate assumption can be investigated numerically. Here, we provide convincing evidence for the validity of the rate assumption.
This lends further support to the idea that generalized Euler-type equations have a structure common to all integrable extensive systems.

\newpage
\section{Introduction, collision rate assumption}
\label{sec1}
\setcounter{equation}{0} 
Euler equations are based on microscopic conservation laws, with possible additions when symmetries are broken. One might wonder
whether such a structure remains valid for one-dimensional integrable many-body systems, for which the number of conservation 
laws is extensive. There has been a lot of interest in this topic, mostly driven by the investigation of quantum integrable systems. We refer to a few fundamental contributions~\cite{Z,CDY16,BCNF16,IN17,BVKM18,IN172,BVKM17,UOKS19,SBDD18}, in which the reader can find a more comprehensive list.

The basic ingredients are easily recalled. For a given model one has to determine all local (perhaps quasi-local) conservation laws. Then one writes 
down the generalized Gibbs ensemble (GGE). In many cases its free energy can be obtained from a Bethe ansatz. The GGE averaged conserved
fields are then first derivatives of this free energy. Determining GGE averaged currents is a more complicated story. In most models it is already difficult to
write down an expression for the microscopic current and computing the GGE average current can be accomplished only in very exceptional cases.  
One thus has to a rely on an educated guess that we call the collision rate assumption, which is first explained in general terms. 

Consider a uniform, one-dimensional fluid, consisting of quasi-particles with bare velocity $v \in  \mathbb{R}$. Quasi-particles of velocity $v$ 
have a uniform spatial density 
$\rho_\mathrm{p}(v)$, that does not change under the dynamics. Two colliding quasi-particles retain their velocity but undergo a spatial shift. Quantum mechanically this would correspond to the two-particle phase shift. When viewed over longer time intervals, 
the intrinsic velocity $v$ of a quasiparticle is modified to an effective velocity $v^\mathrm{eff}(v)$ through collisions. To compute this velocity,
we add a tracer quasi-particle with velocity $v$ at the origin. Upon colliding with a fluid quasi-particle of velocity $w$, the position of the tracer is shifted by $\Phi(v,w)$. Let us now consider a time span $\Delta t $, which is short on the macro-scale but long on the micro-scale. Then $v^\mathrm{eff}(v)\Delta t$
is the true displacement of the tracer particle up to time $\Delta t$. $v\Delta t$ is its bare displacement to which one has to add the 
contribution from collisions. The collision rate is $\rho_\mathrm{p}(w)|v^\mathrm{eff}(w) - v^\mathrm{eff}(v)|$, which leads to 
\begin{eqnarray}\label{1.1}
&&\hspace{-20pt}v^\mathrm{eff}(v) \Delta t 
 = v\Delta t + \Big(\int_v^\infty dw\Phi(v,w) \rho_\mathrm{p}(w)(v^\mathrm{eff}(w) - v^\mathrm{eff}(v))\nonumber\\
&&\hspace{80pt}+ \int_{-\infty}^v dw \Phi(v,w)\rho_\mathrm{p}(w)(v^\mathrm{eff}(v) - v^\mathrm{eff}(w)\Big)\Delta t.
\end{eqnarray}
As a consequence the macroscopic current, used in a hydrodynamic description, is given by $v^\mathrm{eff}(v)\rho_\mathrm{p}(v)$.
 
According to our current understanding the relation \eqref{1.1} holds for \textit{all} one-dimensional integrable many-particle systems,
classical and quantum. There also seems to be no small parameter, except that the conserved fields are assumed to vary slowly on the microscopic scale.
Of course, the labelling of the quasi-particles could be more complicated, and the same is true for the bare velocity.  But $\Phi$ is always deduced from the microscopic two-particle phase shift. So what is the evidence for the collision rate assumption \eqref{1.1}? 

Much work has been done for quantum systems, such as the integrable spin-$1/2$ XXZ chain \cite{BCNF16,IN17,BVKM18,BVKM17,UOKS19}.
One typically starts from a natural initial state, such as a domain wall or a localized cloud in infinite space, and solves numerically the corresponding kinetic equation, which is based on the collision rate assumption. Some specific averages of interest are then compared    
with DMRG simulations of the microscopic dynamics. While the agreement is often excellent, the test as such involves many intermediate steps, although recent analytical work \cite{UOKS19} has demonstrated that the collision rate assumption for averaged spin and energy currents in the XXZ model follows directly from thermodynamic Bethe ansatz (TBA).
On the classical side, an early work discusses the KdV equation \cite{Z71}. One prepares initially a low density of solitons, adds a tracer soliton,
and verifies quite accurately the properly adapted Eq. \eqref{1.1} \cite{CDE16}. 
The regime of high soliton density is addressed in~\cite{E03}, where the collision rate assumption is verified by an analytical argument involving the thermodynamic limit of the Whitham equations (see also ~\cite{EK05}).
For a fluid of hard rods with diameter $a$, Eq. \eqref{1.1} can be easily verified, since the tracer particle collisions turn out to be statistically independent and, by definition of the model, the shift $\Phi = \pm a$ is independent of $v,w$ \cite{DS17a}. The interpretation of the quantum rate assumption in terms of classical flea-gas dynamics was developed in \cite{DTC18, DSY18}, while the connection to the Whitham modulation theory of classical soliton lattices was discussed in \cite{BVKM18,B17}.

There is one further classical system for which \eqref{1.1} has been established recently \cite{FNRW18}. It is a cellular automaton known as the
particle-box system \cite{TS90}. One considers the one-dimensional lattice $\mathbb{Z}$ (= boxes) and each site is either occupied by a particle (= ball) or empty.
Thus a configuration is an infinite string of $0,1$'s. To define a single time step we assume a finite number of particles and introduce a carrier  
which starts to the left with load 0 and moves step-by-step to the right. At the first encounter with a particle, say at site $j$, the site $j$ switches to empty and the  
carrier load increases to $1$. If now site $j+1$ happens to be empty, then the carrier drops its particle, i.e.   the occupation at $j+1$ switches to $1$, while the carrier load decreases to $0$.
Otherwise, if site $j+1$ happens to be occupied, then the carrier picks up this particle, i.e. the occupation at site $j+1$ switches to $0$ and the carrier load increases to $2$,  \textit{etc.}, until the whole sweep 
is completed resulting in the time $1$ configuration of particles and carrier load 0. A $k$-soliton is a configuration of the form $111...000...$ each block of length $k$, $k = 1,2,...$\,.
The bare velocity of the soliton is simply $k$. If a $k$-soliton bumps into a slower $m$ soliton, $m < k$, then the $k$-soliton is shifted  
by $2m$ and the $m$-soliton by $-2m$. The obvious adaptation of \eqref{1.1} is proved for a large class of space-time stationary probability measures on configurations.
The proof is difficult and uses a particular representation of the dynamics, which is hidden behind  the bare definition the box-ball system.
  
In our paper we investigate the collision rate assumption for the classical Toda lattice \cite{S19,D19,BCM19}. This model has several simplifying features.
Firstly, it is classical, so that the GGE can be sampled by Monte-Carlo methods. Secondly, there is a simple and, in a sense, very explicit expression for the microscopic currents. Furthermore, the GGE free energy comes from a variational principle, whose minimizer can be obtained from Dyson's Brownian motion,
which simplifies the task of finding solutions to the TBA equation.
Our goal is to check the collision rate assumption directly, by computing the GGE average of the microscopic currents.
\newpage
\section{Hydrodynamics for non-integrable chains}
\label{sec2}
\setcounter{equation}{0}
We briefly recall the hydrodynamics of a non-integrable classical chain \cite{S14}. 
Considered are $N$ particles in one dimension with positions, $q_j$, momenta $p_j$, $j=1,\ldots,N$. Their mass is set equal to one. Nearest neighbours in index space are coupled through a chain potential $V_\mathrm{ch}$. The Hamiltonian of the chain then reads
\begin{equation}\label{2.1}
  H=\sum^{N}_{j=1} \tfrac{1}{2}p^2_j+\sum^{N-1}_{j=1}V_\mathrm{ch}(q_{j+1}-q_j)\,,
\end{equation}
resulting in the equations of motion
\begin{equation}\label{2.2}
\ddot{q}_j(t)  =  V_\mathrm{ch}'(q_{j+1}(t) - q_{j}(t)) - V_\mathrm{ch}'(q_{j}(t) - q_{j-1}(t))\,.
\end{equation}
We regard \eqref{2.2} as a discrete nonlinear wave equation.  One could also think of particles moving on the real line. Their positions are not ordered and  particles have a somewhat peculiar    
interaction, which is not invariant under relabelling. This second picture is referred to as ``fluid''. As discussed at length in \cite{D19}, one can switch 
back and forth
between the fluid and chain pictures. We adopt here the latter picture, which turns out to be somewhat simpler.

It will be convenient to introduce the positional differences 
\begin{equation}\label{2.3}
r_j=q_{j+1}-q_j\,,
\end{equation}
more physically referred to as the stretch, which can be negative. $r_j$ is the free ``volume'' between 
particles $j+1$ and  $j$. The equations of motion \eqref{2.2} then turn into
\begin{equation}\label{2.4}
\dot{r}_j=p_{j+1}-p_j\,,\qquad
\dot{p}_j=V_\mathrm{ch}'(r_j)-V_\mathrm{ch}'(r_{j-1})\,,
\end{equation}
$j=1,\ldots,N$, where periodic boundary conditions have been adopted, $p_{N+1}=p_1$, $ r_0=r_N$. Viewed as a classical  lattice field theory, the underlying lattice is $[1,\ldots,N]$ with periodic boundary conditions and the field variables are $(r_j,p_j)$. Dynamically, $r_j$ is coupled to its right and $p_j$ to its left neighbour. From the dynamics, one can read off the local conservation laws, namely stretch, momentum, and energy,
\begin{equation}\label{2.5}
\big(r_j, p_j,e_j\big), \qquad e_j=\tfrac{1}{2}p^2_j + V_\mathrm{ch}(r_j),
\end{equation}
and their currents
\begin{equation}\label{2.6}
 \big( -p_j,-V_\mathrm{ch}'(r_{j-1}), - p_jV_\mathrm{ch}'(r_{j-1})\big).
\end{equation}
The assumption of being non-integrable can be rephrased as having no further local conservation laws than already listed.

Hydrodynamics is based on the assumption that local equilibrium is approximately maintained over the course of time evolution. While plausible, such a dynamical property is difficult to deduce 
from the equations of motion. Nevertheless, to write down the correct macroscopic equations one merely has to average the local fields and currents 
in an equilibrium state.
With our choice of variables, the canonical equilibrium states are of product form, with one factor given by 
\begin{equation}\label{2.7}
Z^{-1} \exp\big[-\beta\big(\tfrac{1}{2}(p_0-u)^2 +V_\mathrm{ch}(q_0)+\tilde{P}q_0)\big)\big] \,.
\end{equation}
To yield a well-defined partition function, the chain potential should be bounded from below and  have an at least one-sided, say, linearly  increasing lower bound. The pressure, $\tilde{P}$, which might have to be restricted to a half-line, is dual to the stretch. $\beta$ is the inverse temperature and $u$ the 
mean velocity. Note that
\begin{equation}\label{2.8}
  \tilde{P}=-\langle V'_\mathrm{ch}\rangle_{\tilde{P},\beta}
\end{equation}
is the average force between neighbouring particles. The required thermal averages are easily  accomplished. For the fields, one obtains 
\begin{equation}\label{2.9}
\langle (r_j, p_j,e_j)\rangle_{\tilde{P},u,\beta}= (\ell, u , \mathfrak{e}),\qquad \mathfrak{e} = \tfrac{1}{2}\beta^{-1} +  \langle V_\mathrm{ch}\rangle_{\tilde{P},\beta},
\end{equation}
and their currents,
\begin{equation}\label{2.10}
  \langle (- p_j,-V_\mathrm{ch}'(r_{j-1}) , - 
  p_j V_\mathrm{ch}'(r_{j-1}))\rangle_{\tilde{P},u,\beta} = (-u,\tilde{P}, u\tilde{P}).
\end{equation}

To arrive at an explicitly closed system, one still has to reexpress the average currents in terms of the conserved fields.
In other words, the intensive parameters of the canonical ensemble have to be substituted by the extensive ones, which can always be done due to the convexity of the free energy.
 If we use $x\in\mathbb{R}$ for the corresponding continuum approximation, the conserved fields are the local stretch $\ell(x)$, the local momentum $\mathsf{u}(x)$, and the local total energy $\mathfrak{e}(x) =\frac{1}{2} \mathsf{u}(x)^2+ \mathsf{e}(x)$ per particle, with  
 internal energy $\mathsf{e}$. The microscopic conservation laws then turn into the Euler equations of the nonlinear chain, as
\begin{equation}\label{2.11}
\partial_t\ell +\partial_x \mathsf{j}_\ell =0\,,\quad   \partial_t \mathsf{u} +\partial_x \mathsf{j}_\mathsf{u} =0\,,\quad   \partial_t \mathfrak{e} +\partial_x \mathsf{j}_\mathfrak{e} =0\,,
\end{equation}
where the hydrodynamic currents are given by
\begin{equation}\label{2.12}
(\mathsf{j}_\ell,\mathsf{j}_\mathsf{u},\mathsf{j}_\mathfrak{e})=
 \big(-\mathsf{u},\tilde{P}(\ell,\mathfrak{e}-\tfrac{1}{2}\mathsf{u}^2), \mathsf{u} \tilde{P}(\ell,\mathfrak{e}-\tfrac{1}{2}\mathsf{u}^2)\big).
\end{equation}
Here $\tilde{P}(\ell,\mathsf{e})$ results from inverting the map $(\tilde{P},\beta) \mapsto (\ell,\mathsf{e})$. 

The Toda lattice is a special anharmonic chain with interaction potential
\begin{equation}\label{2.13}
V_\mathrm{ch}(r)= \mathrm{e}^{-r},
\end{equation}
which requires the restriction $\tilde{P} > 0$. With $N$ degrees of freedom, the Toda chain has $N$ independent local conservation laws.
To write down its Euler-type equations, we will still use the blue-print provided by the non-integrable case.

\section{Lax matrix and GGE averaged fields}
\label{sec3}
\setcounter{equation}{0}
The conserved fields and currents of the Toda chain may be obtained concisely from the Lax matrix $L_N$ \cite{F74}.
We define
\begin{equation}\label{3.1}
a_j = \tfrac{1}{2}\mathrm{e}^{-r_j/2},\quad b_j = \tfrac{1}{2} p_j.
\end{equation} 
Then the finite $N$ Lax matrix  is the tridiagonal real symmetric matrix with
\begin{equation}\label{3.2}
(L_N)_{j,j} = b_j, \quad (L_N)_{j,j+1} = (L_N)_{j+1,j}= a_j, 
\end{equation} 
  $(L_N)_{1,N} = (L_N)_{N, 1} = a_N$ because of periodic boundary conditions, and $(L_N)_{i,j}=0$ otherwise. 
For infinite volume, the index is extended to $j \in \mathbb{Z}$ and the corresponding tridiagonal matrix is denote by $L$.
We also introduce $L^\mathrm{off}$ as the off-diagonal part of $L$. 
The conserved fields of the Toda chain are then given by 
\begin{equation}\label{3.3} 
Q^{[n],N} = \mathrm{tr}\big[(L_N)^n\big]  
\end{equation} 
with the obvious density
\begin{equation}\label{3.4} 
Q^{[n]}_j = L_{j,j}. 
\end{equation} 
From the equations of motion one obtains the corresponding total currents \cite{S19},
 \begin{equation}\label{3.5}
J^{[n],N} = \mathrm{tr}\big[(L_N)^nL_N^\mathrm{off}\big]
\end{equation} 
with density
 \begin{equation}\label{3.6}
J^{[n]}_j = (L^nL_\mathrm{off})_{j,j}.
\end{equation} 
As in the non-integrable case, in addition there is the conserved stretch and its current
\begin{equation}\label{3.7} 
Q^{[\mathrm{s}]}_j = r_j, \qquad  J^{[\mathrm{s}]}_j = -p_j.
\end{equation}

The finite volume generalized Gibbs ensemble (GGE) is  given by 
\begin{equation}\label{3.8}
(Z_{\mathrm{toda},N})^{-1}\mathrm{e}^{-\mathrm{tr}[V(L_N)]}\prod_{j=1}^N \mathrm{e}^{ -Pr_j}\mathrm{d}r_j\mathrm{d}p_j, \quad P>0.
\end{equation} 
One should think of the potential $V$ as a finite power series with a strictly positive even leading coefficient. Physically this would correspond to a 
finite number of chemical potentials.  But presumably a larger class of $V$ will work as well. Thermal equilibrium corresponds to 
$V(x) = 2\beta (x- \tfrac{1}{2}u)^2$, compare eq. \eqref{2.7}. However, for notational simplicity we set $P = \beta \tilde{P}$.
The Toda free energy is defined through the infinite volume limit 
\begin{equation}\label{3.9}
F_\mathrm{toda}(P,V)=  -\lim_{N \to \infty} \tfrac{1}{N} \log Z_{\mathrm{toda},N}.
\end{equation}

The Lax matrix has the eigenvalues $\lambda_\ell$ and eigenvectors $\psi_\ell(j)$, both depending on $N$,
\begin{equation}\label{3.10}
L_N \psi_\ell(j) = \lambda_\ell \psi_\ell(j), \quad \ell = 1,...,N.
\end{equation}
The observables of interest are then  
  \begin{eqnarray}\label{3.11}
&&\frac{1}{N} Q^{[n],N} =  \int_\mathbb{R} dx x^n \frac{1}{N}  \sum_{\ell=1}^N \delta (x - \lambda_\ell) x^n,\\\label{3.12}
&&\frac{1}{N} J^{[n],N} = \int_\mathbb{R} dx x^n\frac{1}{N}  \sum_{\ell=1}^N \sum_{j=1}^N 2 a_j\psi_\ell(j) \psi_\ell(j+1) \delta (x - \lambda_\ell).
\end{eqnarray}
The first integrand above is the normalized density of states (DOS) of the Lax matrix with entries distributed according to GGE with parameters $P,V$.
In the second expression there is an extra weight coming from the eigenvectors. In the limit $N \to \infty$ we expect both densities to converge to a limit,
\begin{equation}\label{3.13}
\frac{1}{N}  \sum_{\ell=1}^N  \delta (x - \lambda_\ell) \to  \rho_Q(x), \quad 
 \frac{1}{N}  \sum_{\ell=1}^N \sum_{j=1}^N 2 a_j\psi_\ell(j) \psi_\ell(j+1) \delta (x - \lambda_\ell) \to  \rho_J(x).
\end{equation}
In fact, this limit should be self-averaging. By definition, $ \rho_Q(x) \geq 0$ and $\int dx \rho_Q(x) =1$. But the weights in \eqref{3.12} have no definite sign
and the same is true for  $\rho_J(x)$.
Denoting the finite volume GGE expectation by $\langle \cdot \rangle_{N,P,V}$, and $\langle \cdot \rangle_{P,V}$ its infinite volume limit, one arrives at
 \begin{equation}\label{3.14}
\lim_{N\to \infty} \frac{1}{N} \langle Q^{[n],N} \rangle_{N,P,V} = \langle  (L^n)_{0,0} \rangle_{P,V} =  \int_\mathbb{R} dx \rho_Q(x) x^n
\end{equation}
for the conserved fields and correspondingly for the currents, 
\begin{equation}\label{3.15}
\lim_{N\to \infty} \frac{1}{N} \langle J^{[n],N} \rangle_{N,P,V}   = \langle  (L^nL^\mathrm{off})_{0,0} \rangle_{P,V} = \int_\mathbb{R} dx \rho_J(x) x^n.
\end{equation}
(Existence of the thermodynamic limits on the left-hand side follows by extensivity.) As will be explained, variational type formulas for $F_\mathrm{toda}(P,V)$ and $\rho_Q(x)$ \cite{S19} are available. There is also a prediction for 
 $\rho_J(x)$ based on the collision rate assumption \cite{D19,BCM19}. So our strategy will be to sample the quantities in \eqref{3.13} and to compare them with the theoretical predictions.

\section{Free energy, random matrix model}
\label{sec4}
\setcounter{equation}{0} 

We start from the free energy functional
\begin{equation}\label{4.1}
 \mathcal{F}_P^\mathrm{MF}(\rho) =  \int _\mathbb{R}\mathrm{d}x \rho(x) V(x)      -P \int _\mathbb{R}\mathrm{d}x\int _\mathbb{R}\mathrm{d}y  \log|x-y|  \rho(x)\rho(y) + 
\int _\mathbb{R}\mathrm{d}x \rho(x) \log \rho(x),
\end{equation} 
which has to be minimized under the constraint
\begin{equation}\label{4.2}
\rho(x) \geq 0,\quad  \int _\mathbb{R}\mathrm{d}x \rho(x) =1.
\end{equation}
Since $\mathcal{F}_P^\mathrm{MF}$ is convex, the minimizer is unique and will be denoted by $\rho^*$.
The $P$-dependence can be trivially shifted to the constraint by  setting $\varrho (x) = P\rho(x)$ and  defining
\begin{equation}\label{4.3}
\mathcal{F}(\varrho) =  \int _\mathbb{R}\mathrm{d}x \varrho(x) V(x)      - \int _\mathbb{R}\mathrm{d}x\int _\mathbb{R}\mathrm{d}y   \log|x-y|\varrho(x) \varrho(y) + 
\int _\mathbb{R}\mathrm{d}x \varrho(x) \log \varrho(x)
\end{equation} 
to be minimized under the constraint
\begin{equation}\label{4.4}
\varrho(x) \geq 0,\quad  \int _\mathbb{R}\mathrm{d}x \varrho(x) =P. 
\end{equation}
Clearly, $P \mathcal{F}_P^\mathrm{MF}(P^{-1}\varrho)=  \mathcal{F}(\varrho) -P\log P$ and
$\varrho^*(x) = P\rho^*(x)$.

Introducing the Lagrange multiplier, $\mu$, the Euler-Lagrange equation for $\mathcal{F}$ reads
\begin{equation}\label{4.5} 
  V(x) -  2 \int_\mathbb{R} \mathrm{d}y  \log|x-y| \varrho_\mu(y) +\log \varrho_\mu(x)  - \mu = 0.
 \end{equation}
Following common practice, one writes $\varrho_\mu=  \mathrm{e}^{-\varepsilon}$ with pseudo-energies $\varepsilon$. Then \eqref{4.5} turns into
the central identity
\begin{equation}\label{4.6} 
  \varepsilon(x) =  V(x) - \mu -  2 \int_\mathbb{R} \mathrm{d}y  \log|x-y|  \mathrm{e}^{-\varepsilon(y)}.
 \end{equation}
Originally, in the case $V(x) = x^2$, this equation was obtained from the semiclassical limit of the TBA equation for the quantum Toda chain,
hence the name classical TBA, although the Bethe ansatz appeared nowhere in our discussion. 

For the GGE Toda free energy one finds that 
\begin{equation}\label{4.7} 
 F_\mathrm{toda}(P,V) =  \partial_P \mathcal{F}(\varrho^*(P,V)) -(1+ 2P) \log 2 - 1.
 \end{equation}
Equivalently 
\begin{equation}\label{4.8} 
  F_\mathrm{toda}(P,V) =  \mu(P,V) -(1+ 2P) \log 2.
 \end{equation}

The DOS, $\rho_Q$, is related to the functional derivative  of the Toda free energy w.r.t. $V$. Using \eqref{4.3}, \eqref{4.4} one obtains
\begin{equation}\label{4.9} 
\rho_Q(x) =  \partial_P\varrho^*(x,P). 
\end{equation}
Let us define
\begin{equation}\label{4.10} 
 P =  \int _\mathbb{R}\mathrm{d}x  \varrho_{\mu(P)}(x).
 \end{equation}
Then $\varrho^*(x,P)= \varrho_{\mu(P)}(x)$ and differentiating w.r.t. $P$,
\begin{equation}\label{4.11} 
\rho_Q(x) =  \langle r_0 \rangle_{P,V}\partial_\mu \rho_\mu (x)\big|_{\mu = P}. 
\end{equation}
Finally one can differentiate TBA w.r.t. $\mu$ to obtain 
\begin{equation}\label{4.12} 
\partial_\mu\varrho_\mu(x)= \rho_\mu(x)\big(1 + 2\int _{\mathbb{R}}\mathrm{d}y  \log|x-y|  \partial_\mu\rho_\mu(y)\big), 
\end{equation}
which replaces the $\mu$-differentiation by an integral equation. 

\begin{figure}[hh]
	\centering
	\includegraphics[width=.45\linewidth]{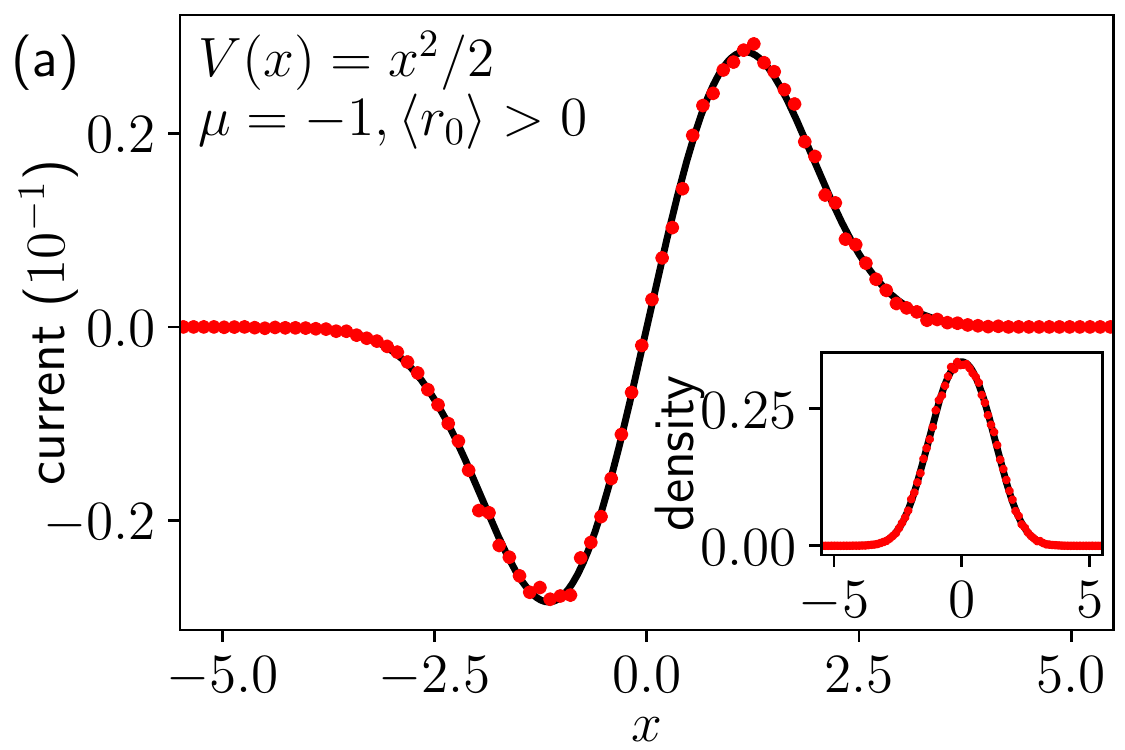}
	\includegraphics[width=.45\linewidth]{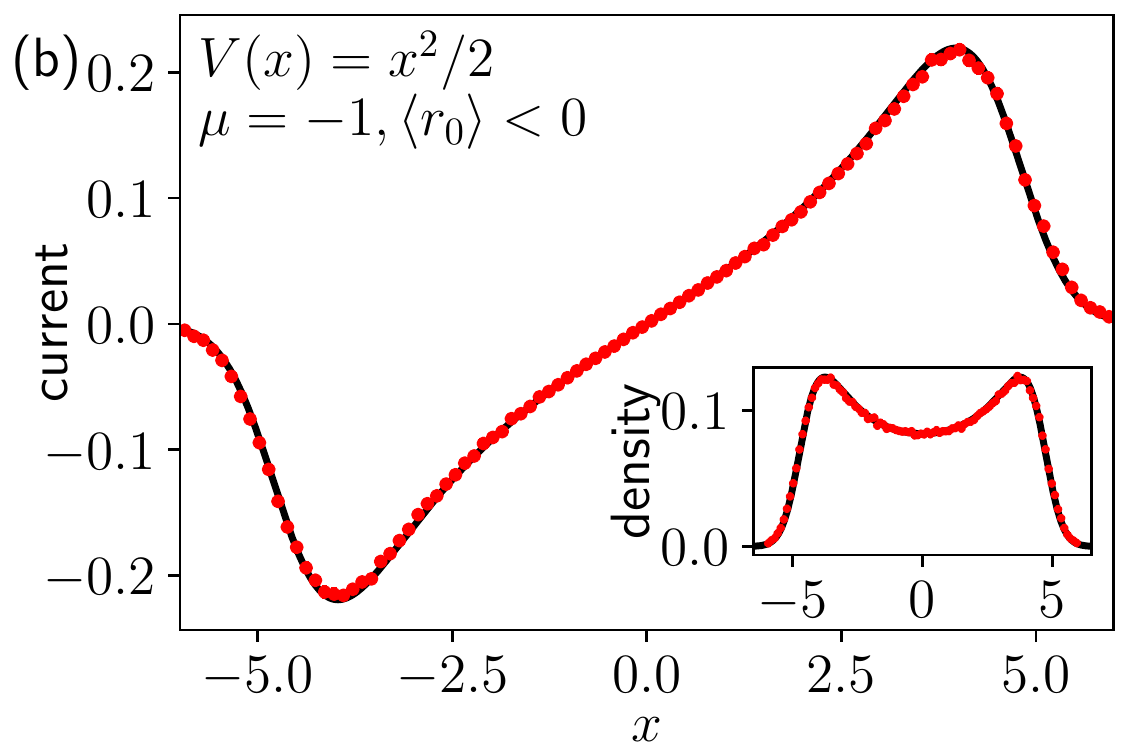} \\
	\includegraphics[width=.45\linewidth]{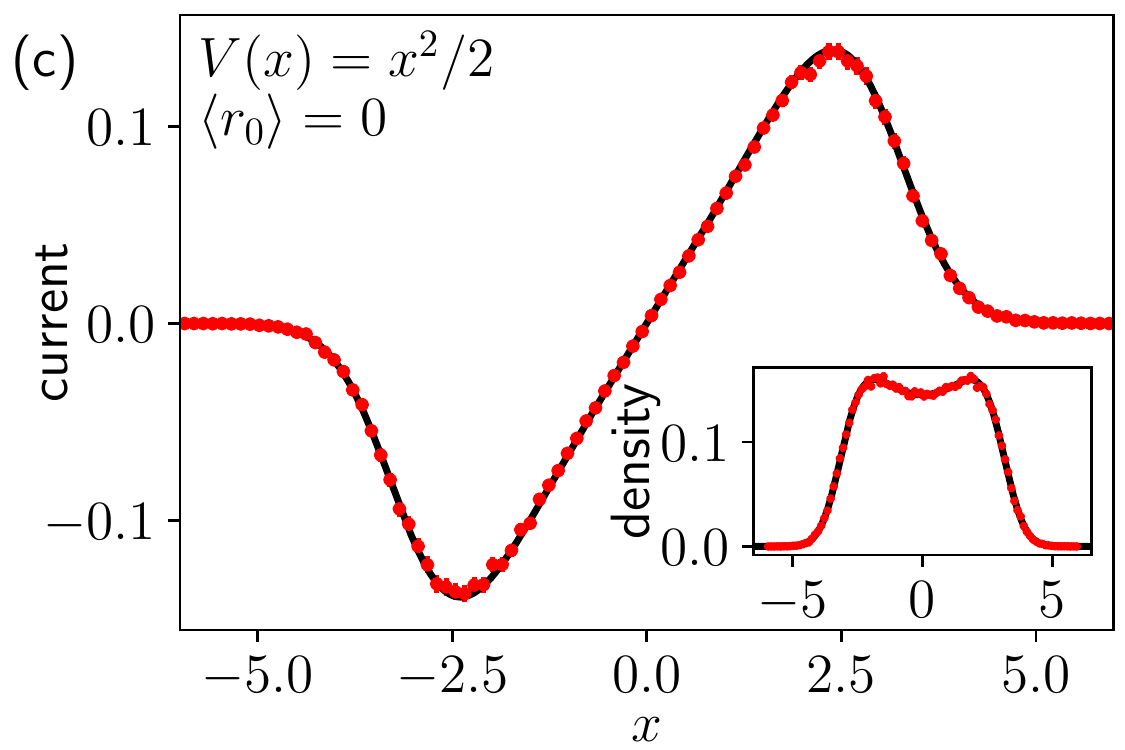}
	\includegraphics[width=.45\linewidth]{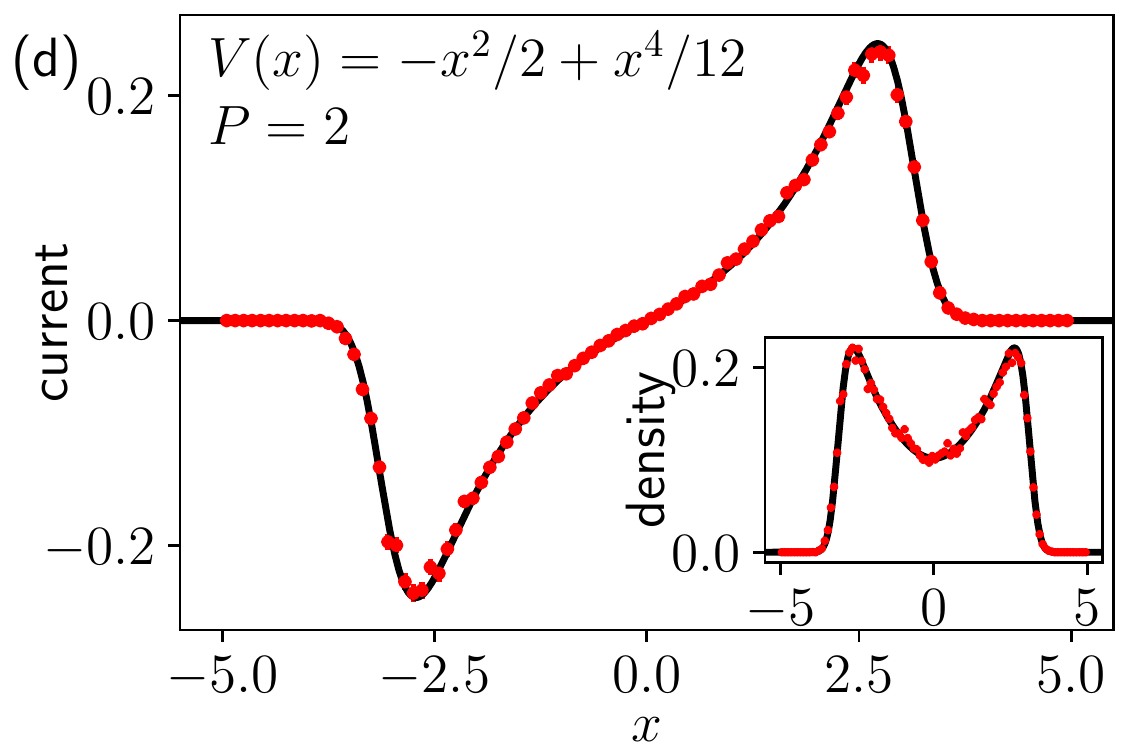}
	\caption{Comparison between the prediction of current (\ref{6.3}) by the collision rate assumption and direct Lax matrix simulation. The insets show the comparison of the density $\rho_Q$, which is a benchmark. The dots are simulation data and the solid curves are obtained from (\ref{6.3}). The errors of both data are smaller than the marker size and line width, respectively. The simulation data for each panel is obtained from averaging over $10^5$ samples of $N=1024$ Lax matrices. (a) and (b) show two solutions to the TBA equation (\ref{4.6}) with the same $V(x)$ and $\mu$. They have $P = 0.138$, $\left<r_0\right>=7.64$  and $P = 4.10$, $\left<r_0\right>=-1.28$ respectively. Panel (c) has $P=1.461$, such that  $\left<r_0\right>=0$. Panel (d) shows an example of non-quadratic potential $V(x)$.  }
	\label{fig1}
\end{figure}

The chemical potential $\mu(P)$ is convex down with maximum $\mu_\mathrm{max}$ at $P_\mathrm{c}$ and $\mu(P)\to -\infty$
for $P \to 1, \infty$. $P < P_\mathrm{c}$ is the low pressure and 
$P > P_\mathrm{c}$ is the high pressure branch. At $P=  P_\mathrm{c}$ one has $\langle r_0 \rangle_{P_\mathrm{c},V} = 0$ and hence $q_N - q_0 = \mathcal{O}(\sqrt{N})$. 
At low pressure the positions are approximately ordered, while at high pressure they are reverse ordered.
Thus for $\mu \neq \mu_\mathrm{max}$ the integral equation (\ref{4.6}) has two solutions \cite{BCM19}. An example is shown in Figure \ref{fig1}.  
 As observed in \cite{BCM19}, the numerical solution of \eqref{4.6} by iteration with fixed $\mu$ converges systematically to the low pressure branch, which should be considered an artifact of the iteration scheme. Instead, the two solutions can be reliably found by a Fokker-Planck or Dyson Brownian motion approach, which  will be outlined below. 
 
For $P \to \infty$, the entropy term in \eqref{4.1} can neglected and one arrives at the much studied variational problem for the ground state of the one-dimensional log gas. For $V(x) = x^2$ its solution is the Wigner semicircle law. Since the Lax matrix DOS is its derivative, one obtains $\rho_Q(x) = 
\pi^{-1}(1- x^2)^{-\frac{1}{2}}$. In Fig. 1, the pressure is chosen to be $P = 4.1$ and one observes that the DOS already starts  to build up a singularity 
at the two edges $x= \pm1$.

\section{Dyson's Brownian motion}
\label{sec5}
\setcounter{equation}{0}

We differentiate TBA w.r.t. to $x$ and obtain
\begin{equation}\label{5.1} 
  V'(x)\rho_{\mu}(x) -  2 P \int_\mathbb{R} \mathrm{d}y \frac{1}{x-y} \rho_{\mu}(x)\rho^*(y) + \partial_x\rho_{\mu}(x) = 0,
 \end{equation}
 which can be viewed as the stationary solution of the nonlinear Fokker-Planck equation
 \begin{equation}\label{5.2}
\partial_t \rho(x,t) = \partial_x\Big( V'(x)\rho(x,t) - 2P \int _\mathbb{R}\mathrm{d}y\frac{1}{x-y}\rho(y,t) \rho(x,t)  + \partial_x  \rho(x,t)\Big).
\end{equation} 
 This suggests that the minimizer can be obtained from simulating many diffusing particles subject to an external potential $V$ and a 
 logarithmic mean field interaction. Their stochastic differential equations read
 \begin{equation}\label{5.3}
dx_j(t) = -V'(x_j(t))dt + \frac{1}{N}\sum_{i = 1,i\neq j}^N \frac{2 P}{x_j(t) - x_i(t)} dt + \sqrt{2} db_j(t), \quad j = 1,...,N, 
\end{equation}
with $\{b_j(t), j = 1,...,N\}$ a collection of independent standard Brownian motions. In the conventional Dyson's Brownian motion
the interaction is strong, i.e. $\mathcal{O}(1)$, and denoted by $\beta$. In this case, the time-stationary distribution is the much studied $\beta$-ensemble of random matrix theory, 
which refers to the energy dominated regime \cite{ABG12}. The variational principle  \eqref{4.1} still holds, provided the entropy term is dropped.  In our context 
the interaction is mean field and entropy and energy are both of order $N$.
Abstractly, it is ensured that the dynamics \eqref{5.3} reaches a unique stationary state and for its  empirical density one has the limiting behavior 
\begin{equation}\label{5.4}
\lim_{N \to \infty} \frac{1}{N} \sum_{j = 1}^N \delta(x-x_j) =  \rho^*(x,P,V).
\end{equation}  
 Thus, rather than to trying to numerically solve TBA, one can run the dynamics \eqref{5.3} until the stationary regime is reached.
 According to \eqref{5.4}, sampling the positions of the particles in the steady state  provides a good approximation to the solution of TBA. 
 
 Alternatively, one may integrate the Fokker-Planck PDE numerically. Starting from an arbitrary initial condition $\rho(x,0)$ such that $\int \rho(x,0) d x = 1$ and $\rho(x,0) > 0$, the solution is guaranteed to converge. In practice, we choose a standard Gaussian initial condition $\rho(x,0) = e^{-x^2/2}/\sqrt{2\pi}$; for all GGE ensembles we considered, convergence (defined in practice by $|\partial_t \rho | < 10^{-5}$ for all $x$) is reached at $t \lesssim 5$.

\section{GGE averaged currents}
\label{sec6}
\setcounter{equation}{0}

For the Toda lattice the collision rate assumption becomes 
\begin{equation}\label{6.1}
v^\mathrm{eff}(x) 
 = x + 2\int_\mathbb{R} \mathrm{d}y \log|x-y| \partial_\mu\rho_\mu(y)\big(v^\mathrm{eff}(y) - v^\mathrm{eff}(x)\big).
\end{equation}
In the case that
\begin{equation}\label{6.2}
\int_\mathbb{R} \mathrm{d}x\rho_Q(x)x = 0,
\end{equation}
the current DOS is conjectured to be equal to
\begin{equation}\label{6.3}
\rho_J(x) = v^\mathrm{eff}(x) \rho_Q(x) /  \langle r_0 \rangle_{P,V}   =  v^\mathrm{eff}(x)  \partial_\mu \rho_\mu(x) \vert_{\mu=P} \,.
\end{equation}

We test this conjecture numerically for a few different GGE states $(P, V)$. For each state, we first integrate the Fokker-Planck equation (\ref{5.2}) until convergence to find the solution to the TBA equation (with a given pressure). The quantities on the RHS of (\ref{6.3}) are obtained by numerically inverting the linear integral equations, see (\ref{4.12}) and also Appendix~\ref{sec7}. This provides the analytical prediction for $\rho_J(x)$, which is to be compared with direct simulations of the Lax matrix (of size $N = 1024$), via Eq.~(\ref{3.15}). When $V(x) = a x^2 / 2$, the matrix elements $a_j$ and $b_j$ are independent and can be sampled directly, see Eq.~(\ref{3.8}). We also consider cases where $V(x)$ is not quadratic. In that case, we sample the GGE ensemble by Markov chain Monte Carlo. In either case, we measure $\rho_J(x)$ using (\ref{3.13}) by exact diagonalization and averaging over many samples. The results for several GGE ensembles are shown in Figure \ref{fig1}. An excellent agreement is found in all tested cases. In particular, we exhibit the two solutions to the TBA equation with a fixed $\mu$, which describe distinct GGE ensembles, with opposite signs of $\langle r_0 \rangle_{P,V}$. The conjecture for the current (\ref{6.3}) appears to be valid for all GGE ensembles. \\\\


\section{Conclusions}
\label{sec7}

While we have tested only four different parameter values, the agreement between theory and numerical results is very convincing, and lends additional support to previous findings. Beyond the rigorously understood case of hard rods \cite{BDS83}, the classical Toda chain appears to be the only integrable many-body system for which the collision rate assumption can be tested directly, for the full current density of states $\rho_J(x)$. 

The dynamical assumption underlying generalized hydrodynamics is the propagation of local GGE states on physically reasonable space-time scales. Our discussion does not touch on this issue at all. Rather, \textit{if} the dynamics of the Toda lattice propagates as a local GGE, then the local Lagrange multipliers have to be updated according to the kinetic equation based on \eqref{6.1}. \\\\
\textit{Acknowledgments.} We thank Benjamin Doyon and Joel E. Moore for helpful comments on the manuscript. We acknowledge support from the ERC synergy Grant UQUAM, the DOE grant DE-SC0019380 (XC) and the DARPA DRINQS program (VBB).

\appendix
\section{Appendix: Some identities}
\label{sec8}
\setcounter{equation}{0}

To be self-contained, we list some well-known facts \cite{CDY16,BCNF16}.
Let us introduce the integral operator
\begin{equation}\label{7.1}
T\psi(x) = 2 \int_\mathbb{R} \mathrm{d}y \log |x-y| \psi(y),\quad x \in \mathbb{R}.
\end{equation}
Then the TBA equations read
\begin{equation}\label{7.2} 
\varepsilon (x)  = V(x)  -  \mu  - (T \mathrm{e}^{-\varepsilon})(x),
 \end{equation}
Setting $\mathrm{e}^{-\varepsilon} = \varrho_\mu$, the dressing of a function $\psi$  is defined by
\begin{equation}\label{7.3} 
\psi^\mathrm{dr} = \psi + T \varrho_\mu \psi^\mathrm{dr},\quad \psi^\mathrm{dr} = \big(1 - T\varrho_\mu\big)^{-1} \psi.
\end{equation}
On the right, $\varrho_\mu$ is regarded as a multiplication operator, $(\varrho_\mu\psi)(x) = \varrho_\mu(x)\psi(x) $.
From \eqref{4.12} we conclude
\begin{equation}\label{7.4} 
\partial_\mu \rho_\mu= (1 - \rho_\mu T)^{-1} \rho_\mu = \rho_\mu(1 - T\rho_\mu)^{-1}[1] = \rho_\mu[1]^\mathrm{dr}.
 \end{equation}
 Here $[1]$ stands for the constant function $\psi(x) = 1$ and similarly $[x]$ for the linear function $\psi(x) = x$.
 
 The collision rate assumption can be expressed as
 \begin{equation}\label{7.5} 
v^\mathrm{eff} = [x] + T\partial_\mu\rho_\mu v^\mathrm{eff} - (T\partial_\mu\rho_\mu)v^\mathrm{eff},
 \end{equation}
 whose solution can be written as
 \begin{equation}\label{7.6} 
v^\mathrm{eff} = \frac{ \,[x]^\mathrm{dr}}{\, [1]^\mathrm{dr}}. 
 \end{equation}
 
To verify we start from \eqref{7.6} and use \eqref{7.4} as 
\begin{equation}\label{7.7} 
 (1 - T \rho_\mu) \frac{\partial_\mu\rho_\mu}{ \rho_\mu} v^\mathrm{eff} = [x],  
 \end{equation}
 equivalently 
 \begin{equation}\label{7.8} 
 \frac {\partial_\mu \rho_\mu}{  \rho_\mu}v^\mathrm{eff} = [x] + T  \partial_\mu \rho_\mu v^\mathrm{eff}.  
 \end{equation}
From \eqref{4.12} we note that the ratio equals $1 + T\partial_\mu \rho_\mu$, thus yielding \eqref{7.5}.

\end{document}